\documentclass[conference]{IEEEtran}
\IEEEoverridecommandlockouts
\usepackage{cite}
\usepackage{amsmath,amssymb,amsfonts}
\usepackage{algorithmic}
\usepackage{graphicx}
\usepackage{textcomp}
\usepackage{xcolor}
\usepackage[export]{adjustbox}
\def\BibTeX{{\rm B\kern-.05em{\sc i\kern-.025em b}\kern-.08em
    T\kern-.1667em\lower.7ex\hbox{E}\kern-.125emX}}
\begin{document}

\title{When SRv6 meets 5G Core: Implementation and Deployment of a Network Service Chaining Function in SmartNICs\\

}

\author{
\IEEEauthorblockN{1\textsuperscript{st} Guilherme Matos}
\IEEEauthorblockA{\textit{Department of Computer Science, UFSCar}\\
Sorocaba, Brazil\\
guilherme.matos@estudante.ufscar.br}\\
\IEEEauthorblockN{3\textsuperscript{rd} Luis Miguel Contreras}
\IEEEauthorblockA{\textit{Telefonica}\\
Madri, Spain\\
luismiguel.contrerasmurillo@telefonica.com}
\and
\IEEEauthorblockN{2\textsuperscript{nd} Leandro C. de Almeida}
\IEEEauthorblockA{\textit{Department of Computer Science, UFSCar}\\
Sorocaba, Brazil\\
leandro.almeida@estudante.ufscar.br}\\
\IEEEauthorblockN{4\textsuperscript{th} Fábio Luciano Verdi}
\IEEEauthorblockA{\textit{Department of Computer Science, UFSCar}\\
Sorocaba, Brazil \\
verdi@ufscar.br}
}

\maketitle

\begin{abstract}

Currently, we have witnessed a myriad of solutions that benefit from programmable hardware. The 5G Core (5GC) can and should also benefit from such paradigm to offload certain functions to the dataplane. In this work, we designed and implemented a P4-based solution for traffic identification and chaining using the Netronome Agilo SmartNIC. The solution here presented is deployed in-between the RAN and UPF (User Plane Function) so that traffic coming from the RAN is identified and chained using SRv6 based on different rules defined by the control plane. The traffic identification and the construction of the SRv6 list of segments are done entirely in the SmartNIC. A minimalist Proof-of-Concept (PoC) was deployed and evaluated to show that this function is perfectly capable to build service function chainings in a transparent and efficient way.
%One of the key technologies of the fifth generation (5G) of mobile communication systems is network slice, which is leveraging network softwarization technologies such as Software Defined Networking (SDN) and Network Functions Virtualization (NFV). In this scenario, is required suport to service fucntion chaining to transforming data processing in a sequence of services. Making use of P4 language and SRv6 protocol we build the Network Service Chaining Function (NSCF), that will work classifing packets and building service function chaining inside the 5G core. Our tests show that this function is perfectly capable to build service function chainings in a transparent and efficient way.
\end{abstract}


\section{Introduction}
\label{cap-introduction}

%The fifth generation (5G) of mobile communication systems is already there, and brings the responsibility to support diversified service requirements that will require a lot of technical and service requirements with respect to throughput, latency, reliability and availability \cite{Rost2017}.

By leveraging network softwarization technologies such as Software Defined Networking (SDN) and Network Functions Virtualization (NFV), a high level of programmability, flexibility, and modularity may be created on top of a common network. 

Aligned with such softwarization, SRv6 has became a key element for the IPv6 data-plane instantiation of Segment Routing \cite{rfc8402}. SRv6 works as an extension of IPv6 header, creating a segment list of IPv6 addressees having a pointer to identify which segment is active. Every time that the packet pass through a segment endpoint (SR-capable nodes whose address is in the IPv6 destination address) the pointer decreases, and the new segment-id of the segment list is copied to the destination address. Undoubtley, SRv6 is an enabler to satisfy new consumer, service and business demands for 5G and beyound \cite{gramaglia2020experimenting}.

In this demonstration, we will show a P4-based solution capable of identifying traffic and building the list of SRv6 segments in the dataplane, named INCA (In-Network IdentifiCation and chAining). The solution is deployed in-between the RAN and UPF so that traffic is transparently captured, identified and chained according to the control plane pre-defined policies. INCA is capable of parsing traffic coming from the RAN, analysing different types of fields for classification such as IPv6 header (inner/outter), TEID (Tunnel Endpoint ID), QoS ID, among others. In this work, GTP is used as the tunneling protocol between the RAN and UPF.

The deployment was done using a Netronome Agilio CX 2x10GbE SmartNIC. We evaluated the solution by using DASH traffic as well as ICMP traffic from an emulated UE to test different service chaining based on general policies. The evaluation shows the feasibility of INCA in supporting such approach entirely in the dataplane.

\section{Design and working flow}
\label{cap-desing and implementation}

\begin{figure*}[h]
    \centering
    \includegraphics[scale=0.58]{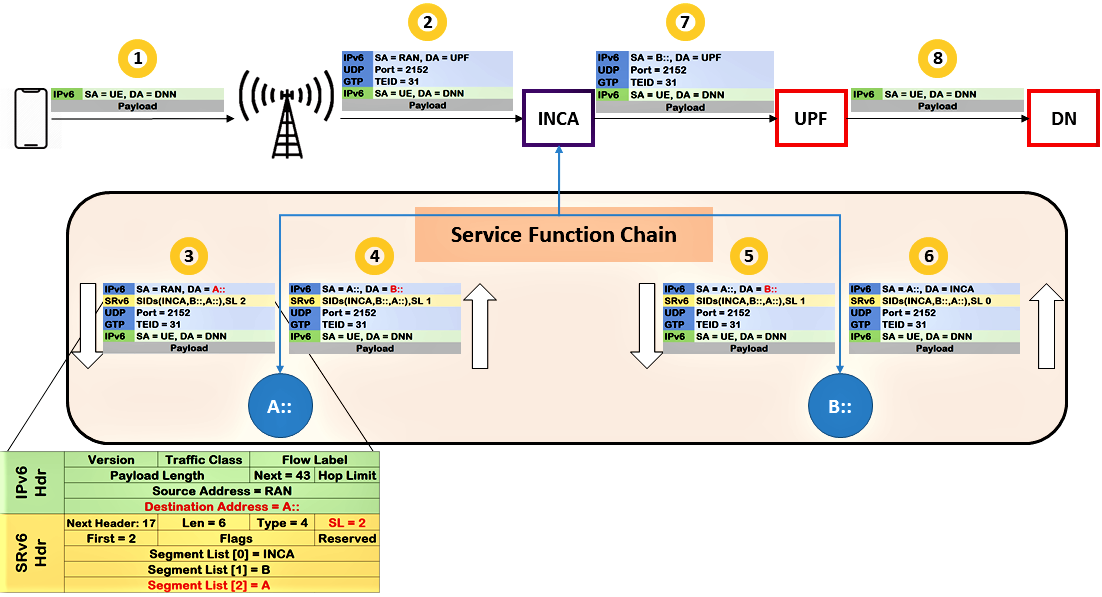}
    \caption{INCA working flow.}
    \label{fig:inca}
\end{figure*}

%This work proposes to make use of the P4 language to build a function that manipulates packets between the RAN and UPF, implementing SRv6 as the data plane protocol to work as a service function chaining inside of 5G architecture.

Figure \ref{fig:inca} ilustrates how our solution works. For sake of simplicity, only the UE, RAN, UPF and DN are shown in the figure. In addition, we also have two examples of network functions, A:: and B::. 
%INCA was implemented to perform the chaining between these functions. 

In (1) a packet leaves the UE to the DN. When this packet arrives in the RAN (2) it is then tunneled with the standard 5G stack (UDP + IPv6 + GTP) and sent to UPF. INCA transparently captures this traffic before the UPF and applies the rules according to what was configured by the control plane. Several fields may be used to control the traffic such as the Tunnel Endpoint ID (TEID - inside GTP), QoS ID, transport and network layers of the user's original packet and slice ID. In addition, we can also detect traffic at the flow level (5-tuple), services, QoS or any combination of those elements.  

In step (3), INCA builds an SRv6 header and forwards the packet to the first function. The last VNF forwards the packet back to INCA so that the SRv6 header is removed and the original traffic is sent to the UPF (7), which in turn delivers the packet to its destination (8).e

%This allows for high and low granularity, where in the same slice there can be different functions in which the chaining itself is carried out in the desired traffic. Thus, in (3) INCA builds an SRv6 header and forwards the packet to the first function. After the packet traverse all defined functions, INCA then removes the SRv6 header and sends it to the UPF (7), which in turn delivers the package to its destination (8).
 
%Taking in account that SRv6 cannot be used as 5G data plane transport protocol since other functions still depend on the GTP, INCA is in charge of creating and removing the SRv6 header, in addition to forwarding the packets to the functions in a totally independent and transparent way. Thus, we can enable a service function chaining inside the 5G core without the need to change the original transport protocol or others predefined functions.

\section{PoC deployment}
\label{cap-deployment and evaluation}

The Netronome SmartNIC uses single-root input/output virtualization (SRIOV), which enables virtual functions (VFs) to be created from a physical function (PF). The VFs thus share the resources of a PF, while VFs remain isolated from each other. The isolated VFs are typically assigned to virtual machines (VMs) on the host. In this way, the VFs allow the VMs to directly access the PCI device, thereby bypassing the host kernel. In our solution, we have two physical (p0, p1) and five virtual interfaces (Vf0\_1 to Vf0\_5).

\begin{figure}[!ht]
    \centering
    \includegraphics[width=8cm]{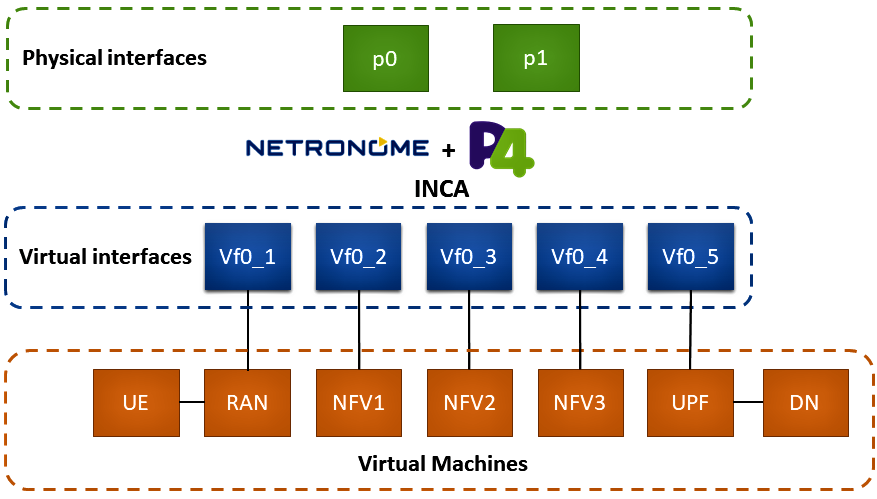}
    \caption{Testbed setup.}
    \label{fig:netronome}
\end{figure}

Figure \ref{fig:netronome} summarizes the setup, showing seven virtual machines used to host the UE, RAN, UPF and DN, as well as three virtual functions: NFV1 (Intrusion Detection System), NFV2 (Intrustion Prevention System) and NFV3 (Packet Filter). To encapsulate and decapsulate packets into GTP tunnels (in the RAN and UPF) we are using Python scripts with the Scapy library.

In this scenario, the UE runs two applications: a DASH client (VLC) and a ICMP monitoring tool. Using the QoS ID field, the DASH traffic is identified to go through two virtual functions, NFV1 and NFV2. The ICMP traffic is identified to gothrough the functions NFV1 and NFV3. This last function is configured to block ICMP traffic.

No scalability evaluation was done so far. The PoC here presented shows that it is possible to build SFC using SRv6 entirely in the dataplane. The NFs and the services used are just examples of what can be done once the INCA is running, and any other NF can be used since INCA is agnostic of the functions applied in the traffic. 

In summary, INCA may be used as a starting point framework to create several different chainings using SRv6 in the dataplane. In addition, the usage of Stratum \cite{stratum} for configuring the rules in INCA is a natural step in this work so that INCA becomes adherent to the ONF next generation SDN.

%In this way we can show INCA classifying two differents traffics based on differents attributes, besides of building two different chains.

%\input{4Conclusion}
\bibliographystyle{IEEEtran} 
\bibliography{references.bib}

\end{document}